\documentclass[submission,copyright,creativecommons]{eptcs}
\providecommand{\event}{FICS 2012} 
\usepackage{breakurl}             

\title{Model-Checking the Higher-Dimensional Modal $\mu$-calculus}
\author{Martin Lange \qquad Etienne Lozes 
\institute{School of Electr.\ Eng.\ and Computer Science, University of Kassel, Germany}}
\def\titlerunning{Model-Checking the Higher-Dim.\ $\mu$-Calculus}
\def\authorrunning{M.~Lange, E.~Lozes}

\usepackage{hmodal}

\newcommand{\Mudiam}[2]{\ensuremath{\langle #1 \rangle_{#2}}}
\newcommand{\Mubox}[2]{\ensuremath{[ #1 ]_{#2}}}

\newcommand{\subf}[1]{\ensuremath{\mathit{Sub}(#1)}}
\newcommand{\size}[1]{\ensuremath{|#1|}}

\newcommand{\prdc}[1]{\mathsf{#1}}
\newcommand{\vtwo}{\ensuremath{\mathscr{V}}}

\begin{document}

\maketitle

\begin{abstract}
  The higher-dimensional modal $\mu$-calculus is an extension of the $\mu$-calculus in which formulas
  are interpreted in tuples of states of a labeled transition system. Every property that can be
  expressed in this logic can be checked in polynomial time, and conversely every polynomial-time
  decidable problem that has a bisimulation-invariant encoding into labeled transition systems can also
  be defined in the higher-dimensional modal $\mu$-calculus.  We exemplify the latter connection by
  giving several examples of decision problems which reduce to model checking of the higher-dimensional
  modal $\mu$-calculus for some fixed formulas. This way generic model checking algorithms for the logic
  can then be used via partial evaluation in order to obtain algorithms for theses problems which may
  benefit from improvements that are well-established in the field of program verification, namely
  on-the-fly and symbolic techniques. The aim of this work is to extend such techniques to other fields
  as well, here exemplarily done for process equivalences, automata theory, parsing, string problems,
  and games.
\end{abstract}

\section{Introduction}

The Modal $\mu$-Calculus $\hmu{}$ \cite{Kozen83} is mostly known as a
backbone for temporal logics used in program specification and
verification. The most important decision problem in this domain is the
model checking problem which is used to automatically prove correctness
of programs. The model checking problem for $\hmu{}$ is well-understood
by now. There are several algorithms and implementations for it. It is
known that model checking $\hmu{}$ is equivalent under linear-time
translations to the problem of solving a parity game \cite{Stirling95}
for which there also is a multitude of algorithms available. From a
purely theoretical point of view, there is still the intriguing question
of the exact computational complexity of model checking $\hmu{}$: the
best known upper bound for finite models is UP$\cap$coUP
\cite{Jurdzinski/98}, which is not entirely matched by the P-hardness
inherited from model checking modal logic.

$\hmu{}$ can express exactly the bisimulation-invariant properties of 
tree or graph models which are definable in Monadic Second-Order Logic \cite{CONCUR::JaninW1996},
i.e.\ are regular. This means that for every such set $L$ of trees or graphs there 
is a fixed $\hmu{}$ formula $\varphi_L$ s.t.\ a tree or
graph $G$ is a model of $\varphi_L$ iff it belongs to $L$. Thus, any decision
problem that has an encoding into regular and bisimulation-invariant sets of trees or graphs
can in principle be solved using model checking technology. In detail, suppose
there is a set $M$ and a function $f$ from the domain of $M$ to graphs s.t.\ 
$\{ f(x) \mid x \in M \}$ is regular and
closed under bisimilarity. By the result above there is an $\hmu{}$ formula
$\varphi_M$ which defines (the encoding of) $M$. Now any model checking algorithm
for $\hmu{}$ can be used in order to solve $M$.

Note that in theory this is just a reduction from $M$ to the model checking problem
for $\hmu{}$ on a fixed formula. Obviously reductions from any problem $A$ to some
problem $B$ can be used to transfer algorithms from $B$ to $A$, and the algorithm
obtained for $A$ can in general be at most as good as the algorithm for $B$ unless
it can be optimised for the fragment of $B$ resulting from embedding $A$ into it.
However, there are two aspects that are worth noting in this context.
\begin{itemize}
\item A reduction to model checking for a fixed formula can lead to much more efficient
      algorithms. A model checking algorithm takes two inputs in general: a structure
      and a formula. If the formula is fixed then partial evaluation can be used in
      order to optimise the general scheme, throw away data structures, etc.
\item Program verification is a very active research area which has developed many
      clever techniques for evaluating formulas in certain structures including 
      on-the-fly \cite{Stirling95} and symbolic methods \cite{IC::BurchCMDH1992}, 
      partial-order reductions, etc.
\end{itemize}
We refer to \cite{al-lpar07} for an example of this scheme of reductions to model
checking for fixed formulas, there being done for problems that are at least 
PSPACE-hard. It also shows how this can be used to solve computation problems in this
way. Since the data complexity (model checking with fixed formula) of 
$\hmu{}$ is in P, using this scheme for $\hmu{}$ is restricted to computationally
simpler problems which can nevertheless benefit from developments in program
verification. Furthermore, it is the presence of fixpoint operators in such a logic
which makes it viable to this approach: fixpoint operators can be used to express
inductive concepts---e.g.\ the derivation relation in a context-free grammar---and
at the same time provide the foundation for algorithmic solutions via fixpoint
iteration for instance. 

Here we consider an extension of $\hmu{}$, the Higher-Dimensional Modal $\mu$-Calculus
$\hmu{\omega}$, and investigate its usefulness regarding the possibility to obtain
algorithmic solutions to various decision or computation problems which may benefit
from techniques originally developed for program verification purposes only. It is known
that $\hmu{\omega}$ captures the bisimulation-invariant fragment of P. We will sketch
how the $\hmu{\omega}$ model checking problem can be reduced to $\hmu{}$ model checking 
via a simple product construction on transition systems. Thus we can obtain---in
principle---an algorithm for every problem that admits a polynomial-time solution and
a bisimulation-invariant encoding into graphs. The reduction from $\hmu{\omega}$ to
$\hmu{}$ is compatible with on-the-fly or BDD-based model checking techniques, thus 
transferring such algorithms from $\hmu{}$ first to $\hmu{\omega}$ and then on to
such decision problems. 


\section{The Higher-Dimensional Modal $\mu$-Calculus}
\label{sec:prel}

\paragraph*{Labeled Transition Systems.}
A labeled transition system (LTS) is a graph whose vertices and edges 
are labeled with sets of
propositional variables and labels respectively. Formally, an LTS over a set $\labels =\{a,b,\ldots\}$ of
edge labels and a set $\prop=\{p,q,\ldots\}$ of atomic propositions is a tuple
$\lts=(S,s_0,@D,@r)$ such that $s_0\in S$, $@D\subseteq S\times\labels\times S$ and 
$@r:S->\Pset(\prop)$. Elements of $S$ are called states, and we write 
$\trans s a {s'}$ when $(s,a,s')\in @D$. The state $s_0\in S$ is called
the initial state of $\lts$.

We will mainly consider \emph{finite} transition systems, \emph{i.e.} 
transition
systems $(S,s_0,@D,@r)$ such that $S$ is a finite set. Infinite-state 
transition systems arising from program verification 
are also of interest, but
their model checking techniques differ from the ones of finite LTS and cannot
be handled by our approach (see more comments on that point in the conclusion).

\paragraph*{Syntax.}
We assume infinite sets $\var=\{x,y,\dots\}$ and $\vartwo=\{X,Y,\dots\}$,
of first-order and second-order variables respectively. For tuples of first-order 
variables $\bar x = (x_1,\dots,x_n)$ and $\bar y = (y_1,\dots, y_n)$, 
with all $x_i$ distinct, 
$\bar x <- \bar y$, denotes the function $\kappa:\var->\var$ such that 
$\kappa(x_i)=y_i$, and $\kappa(z)=z$ otherwise. It is called a \emph{variable
replacement}.

The syntax of the higher-dimensional modal $\mu$-calculus $\hmu{@w}$ is 
reminiscent of that of the
ordinary modal $\mu$-calculus. However, modalities and propositions
are relativized to a first-order variable, and it also features the 
\emph{replacement} modality $\repl{\kappa}$. Formulas of $\hmu{\omega}$ are
defined by the grammar
$$
\varphi,\psi \enspace ::= \enspace p(x) \mid X \mid \neg\varphi \mid \varphi \wedge \psi \mid
\Mudiam{a}{x}\varphi \mid \mu X.\varphi \mid \repl{\bar x <- \bar y}\varphi
$$
where $x,y \in \var$, $\kappa:\var->\var$ is a variable replacement with finite
domain, $a \in \labels$, and $X \in \vartwo$. We require that every second-order variable 
gets bound by a fixpoint quantifier $\mu$ at most once in a formula. Then for every 
formula $\varphi$ there is a function $\mathit{fp}_\varphi$ which maps each second-order variable $X$
occurring in $\varphi$ to its unique binding formula $\mathit{fp}_\varphi(X)=\mu X.\psi$. 
Finally, we allow occurrences of a second-order variable $X$ only 
under the scope of an even number of negation symbols underneath $\mathit{fp}_{\varphi}(X)$.

A formula is of dimension $n$ if it contains at most $n$ distinct first-order
variables; we write $\hmu{n}$ to denote the set of formulas of
dimension $n$. Note that $\hmu{1}$ is equivalent to the standard modal 
$\mu$-calulus: with a single first-order variable $x$, we
have $p(x) \equiv p$, $\repl{x<-x}\psi \equiv \psi$ and 
$\Mudiam{a}{x}\psi \equiv \Mudiam{a}{}\psi$
for any $\psi$.
 
As usual, we write $\varphi\vee\psi$, $\Mubox{a}{x}\varphi$, and $\nu X. @f$ to denote 
$\neg (\neg \varphi \wedge \neg\psi)$, $\neg \Mudiam{a}{x} \neg \varphi$, 
$\neg \mu X.\neg \varphi'$ respectively where $\varphi'$ is obtained from $\varphi$ by replacing
every occurrence of $X$ with $\neg X$. Other Boolean operators like $=>$ and $<=>$ are defined
as usual.

Note that $\repl{\kappa}$ is an operator in the syntax of the logic; 
it does not describe 
syntactic replacement of variables. 
Consider for instance the formula
$$
\nu X . \bigwedge_{p\in\prop} p(x) => p(y) ~~ /|~~ 
\bigwedge_{a\in \labels} \Mubox{a}{x}\Mudiam{a}{y} X~~/|~~ \repl{(x,y)<-(y,x)}X.
$$
As we will later see, this formula characterizes bisimilar states $x$ and $y$.
In this formula, the operational meaning of 
$ \repl{x,y<-y,x}X$ can be thought as ``swapping the players' pebbles'' in the
bisimulation game.

We will sometimes require formulas to be in \emph{positive normal form}. Such 
formulas are built from literals $p(x)$, $\neg p(x)$ and second-order 
variables $X$ using the operators $\wedge$, $\vee$,
$\Mudiam{a}{x}$, $\Mubox{a}{x}$, $\mu$, $\nu$, and $\repl{\kappa}$.
A formula is \emph{closed} if all second-order variables are bound by some 
$\mu$.

With $\subf{\varphi}$ we denote that set of all \emph{subformulas} of 
$\varphi$. It also serves as a good measure for the \emph{size} of a 
formula: $\size{\varphi} := \size{\subf{\varphi}}$. Another good measure of 
the complexity of the formula $\varphi$ 
is its \emph{alternation depth} $ad_{\varphi}$, \emph{i.e} the maximal alternation
of $\mu$ and $\nu$ quantifiers along any path in the syntactic tree of 
its positive normal form.

\paragraph*{Semantics.}
A first-order valuation $v$ over a LTS $\lts$ is a mapping from
first-order variables to states, and a second order valuation is a mapping
from second order variables to sets of first-order valuations:
$$
\begin{array}{rclcl}
\set {Val}  & == & \var & -> & S\\
\set {Val}_2 & == & \vartwo & -> & \Pset(\set {Val})
\end{array}
$$

We write $v[\bar x|->\bar s]$ to denote the first-order valuation that 
coincides with $v$, except that 
$x_i\in \bar x$ is mapped to the corresponding $s_i\in \bar s$. 
We use the same notation $\vtwo[\bar X|->\bar P]$ for second-order 
valuations.
The semantics of a formula $\varphi$ of $\hmu{@w}$ for a LTS $\lts$ 
and a second-order valuation $\vtwo$ is defined as a set of first-order
valuations by induction on the formula:
$$
\begin{array}{lcl}
[|p(x)|]_{\lts}^{\vtwo} & \enspace == \enspace & \{v:p\in @r(v(x))\} \\
{[|} \neg \varphi  |]_{\lts}^{\vtwo} & == & \set{Val} - [| \varphi |]_{\lts}^{\vtwo} \\
{[|}\varphi/|\psi|]_{\lts}^\vtwo & == & [|\varphi|]_{\lts}^\vtwo\cap [|\psi|]_{\lts}^\vtwo \\
{[|}\Mudiam{a}{x}\varphi|]_{\lts}^\vtwo & == & \{v:\exists s.
~\trans {v(x)}{a}s~\mbox{and}~v[x|->s]\in [|\varphi|]_{\lts}^\vtwo\} \\
{[|}X|]_{\lts}^\vtwo & == & \vtwo(X)\\
{[|}\mu X.\varphi|]_{\lts}^\vtwo & == & LFP~~ \lambda P\in \Pset(\set{Val}) .~[|\varphi|]_{\lts}^{\vtwo[X|->P]}
\\
{[|}\repl{\bar x<-\bar y}\varphi |]_{\lts}^\vtwo & == & \{v: v[\bar x|->v(\bar y)]\in [|\varphi|]_{\lts}^\vtwo\}
\end{array}
$$

We simply write $[|\varphi|]_{\lts}$ to denote the semantics of a 
closed formula. We write 
$\lts,v|=\varphi$ if $v\in[|\varphi|]_{\lts}$, and $\lts|=\varphi$ if
$\lts,v_0|=\varphi$, where $v_0$ is the constant function to $s_0$. 
Two formulas are equivalent, written $\varphi \equiv \psi$, 
if $[|\varphi|]_{\lts} = [|\psi|]_{\lts}$ for
any LTS $\lts$. As with the normal modal $\mu$-calculus, it is a 
simple exercise to prove
that every formula is equivalent to one in positive normal form.

\begin{proposition}
\label{prop:pnf}
For every $\varphi \in \hmu{\omega}$ there is a $\psi$ in positive normal form such that
$\varphi \equiv \psi$ and $\size{\psi} \le 2\cdot\size{\varphi}$.
\end{proposition}

\paragraph*{Reduction to the Ordinary $\mu$-Calculus.}
Here we consider $\hmu{\omega}$ as a formal language for defining decision problems. 
Algorithms for these problems can be obtained from model checking algorithms for $\hmu{}$ 
on fixed formulas using partial evaluation. In order to lift all sorts of special 
techniques which have been developed for model checking in the area of program verification
we show how to reduce the $\hmu{\omega}$ model checking problem to that of $\hmu{1}$, i.e.\ 
the ordinary $\mu$-calculus.

Let us assume a fixed non-empty finite subset $V$ of first-order variables.
A formula $\varphi$ of $\hmu{\omega}$ with 
$\fv(\varphi)\subseteq V$ can be seen as
a formula $\hat{\varphi}$ of $\hmu{1}$ over the set of the atomic propositions 
$\prop \times V$ and the action labels
$\labels \times V \cup (V \to V)$. We write $p_x$ instead of 
$(p,x)$ for elements of $\prop \times V$, and equally $a_x$ for elements 
from $\labels \times V$. Then $\varphi|->\widehat{\varphi}$ 
can be defined as the
homomorphism such that $\widehat{p(x)} == p_x$,
$\widehat{\Mudiam{a}{x}{\varphi}} == \Mudiam{a_x}{}\widehat{\varphi}$,
and $\widehat{\repl{\bar x<-\bar y}\varphi} == 
\Mudiam{\bar x <- \bar y}{}\widehat{\varphi}$.

We call an LTS \emph{higher-dimensional} when it interprets the
extended propositions $p_x$ and 
modalities $\Mudiam{a_x}{}$ and $\Mudiam{\kappa}{}$
introduced by the formulas $\widehat{\varphi}$, and \emph{ground} when
it interprets the standard propositions and modalities. 
For a ground LTS $\lts$ and a formula $\varphi$, we thus need to define
the higher-dimensional LTS over which $\widehat{\varphi}$ should be
interpreted: we call it the $V$-\emph{clone} 
of $\lts$, and write it $\clone{V}{\lts}$. Roughly speaking, 
$\clone{V}{\lts}$ is the asynchronous product of $|V|$ copies of 
$\lts$. More formally, assume $\lts = (S,s_0,\Delta,\rho)$; then 
$\clone{V}{\lts}=(S',s_0',\Delta',\rho')$ is defined as follows.
\begin{itemize}
\item The states are valuations of the variables in $V$ by states in $S$, \emph{e.g} $S'=V \to S$, and $s_0'$ is the constant function $\lambda x\in V.s_0$.
\item The atomic proposition $p_x$ is true in those new states, 
which assign $x$ to an original state that satisfies
$p$, e.g.\ $\rho'(v)=\{p_x~:~p\in\rho(v(x))\}$.
\item The transitions contain labels of two kinds. 
First, there is an $a_x$-edge between two valuations $v$ and $v'$, if there 
is an $a$-edge between $v(x)$ and $v'(x)$ in the original LTS $\lts$:
\begin{displaymath}
\trans{v}{a_x}{v'} \quad \text{iff} \quad \exists t. \trans{v(x)}{a}{t} \text{ and } v'=v[x|->t]. \\
\end{displaymath}
For the other kind of transitions we need to declare the effect of 
applying a replacement to a 
valuation. Let $v:V->S$ be a valuation of the first-order variables in $V$, 
and $\kappa:V->V$ be a replacement operator. 
Let $\dual{\kappa}(v)$ be the 
valuation such that $\dual{\kappa}(v)(x)=v(\kappa(x))$.
Then we add the 
following transitions to $\Delta'$.
\begin{displaymath}
\trans{v}{\kappa}{v'} \quad \text{iff} \quad v'= \dual{\kappa}(v)
\end{displaymath}
\end{itemize}

Note that the relation with label $\kappa$ is functional for any such 
$\kappa$, i.e.\ every state
in $\clone{V}{\lts}$ has exactly one $\kappa$-successor. Hence, we have 
$\Mudiam{\kappa}{}\psi \equiv \Mubox{\kappa}{}\psi$
over cloned LTS. 

\begin{theorem}
\label{thm:reduction}
Let $V$ be a finite set of first-order variables,
let $\lts = (S,s_0,\Delta,\rho)$ be a ground LTS,  and
let $\varphi$ be a $\hmu{\omega}$ formula such that $\fv(\varphi)\subseteq V$.
Then 
$$
\lts \models \varphi\quad\text{iff}\quad
\clone{V}{\lts}\models \widehat{\varphi}.$$ 
\end{theorem}

The proof goes by straightforward induction on $\varphi$ and is therefore ommitted -- see also the chapter on descriptive complexity 
in~\cite{Gradel_07_finite} for similar results. The
importance of Thm.~\ref{thm:reduction} is based on the fact that it transfers many model
checking algorithms for the modal $\mu$-calculus to $\hmu{1}$, for example on-the-fly
model checking \cite{Stirling95}, symbolic model checking \cite{IC::BurchCMDH1992} with BDDs or via SAT, 
strategy improvement schemes \cite{conf/cav/VogeJ00}, etc.


\section{Various Problems as Model Checking Problems}
\label{sec:express}

The model checking algorithms we mentioned can be exploited to solve any 
polynomial-time problem that can be encoded as a model checking problem
in $\hmu{\omega}$. By means of examples, we now intend to show that
these problems are quite numerous.

\paragraph*{Process Equivalences.}

The first examples are process equivalences encountered in process
algebras.
We only consider here strong simulation equivalence and bisimilarity, 
and let the interested reader think about how to encode 
other process equivalences, like weak bisimilarity for instance.

Let us first recall some standard definitions.
Let $\lts=(S,s_0,@D,@r)$ be a fixed LTS. A 
\emph{simulation} is a binary relation
$R\subseteq S\times S$ such that for all $(s_1,s_2)$ in $R$,
\begin{itemize}
\item for all $p\in\prop$: $p\in @r(s_1)$ iff $p\in @r(s_2)$;
\item for all $a\in \labels$ and $s_1'\in S$, if $\trans {s_1} a {s_1'}$, 
then there is $s_{2}'\in S$ such that $\trans {s_{2}} a {s_{2}'}$ and
$(s_1',s_2')\in R$.
\end{itemize}
Two states $s,s'$ are \emph{simulation equivalent}, $s\simequiv s'$, 
if there are simulations $R,R'$ such that
$(s,s')\in R $ and $(s',s)\in R'$. A simulation $R$ is a 
\emph{bisimulation} if $R=R^{-1}$; we say 
that $s,s'$ are \emph{bisimilar}, 
$s\bisim s'$, if there is a 
bisimulation that contains $(s,s')$. We say that two valuations are 
bisimilar, $v\sim v'$, if for all $x\in \var$, $v(x)\sim v'(x)$.

\begin{proposition}\cite{Otto99}
$\hmu{\omega}$ is closed under bisimulation: if $v\in[|@f|]$ and $v\sim v'$, then
$v'\in [|@f|]$.
\end{proposition}

Let us now explain how these process equivalences can be decided by the
model checking algorithms:
the following formula captures valuations $v$ such that $v(x)\bisim v(y)$
$$
\nu X . \bigwedge_{p\in\prop} p(x) <=> p(y) ~~ /|~~ 
\bigwedge_{a\in \labels} \Mubox{a}{x}\Mudiam{a}{y} X~~/|~~ \repl{(x,y)<-(y,x)}X
$$
whereas the following formula captures valuations $v$ such that 
$v(x)\simequiv v(y)$
$$
\nu X\big(\nu Y . \bigwedge_{p\in\prop} p(x)<=> p(y) ~~ /|~~ 
\bigwedge_{a\in \labels} \Mubox{a}{x}\Mudiam{a}{y}Y \big)~/|~\repl{(x,y)<-(y,x)}X.
$$

\paragraph*{Automata Theory.}

A second application of $\hmu{\omega}$ is  in the field of automata theory.
To illustrate this aspect, we pick some language inclusion problems that
can be solved in polynomial-time.

A non-deterministic B\"{u}chi automaton can be viewed as a finite 
LTS $A=(S,s_0,\Delta,\rho)$ where $\rho$ interprets
a predicate $\prdc{final}$. Remember that a run on an infinite word 
$w\in\Sigma^\omega$ in $A$ is accepting if it visits infinitely often a
final state. The set of words $L(A)\subseteq \Sigma^\omega$ that have
an accepting run is called the language accepted by $A$. 

The language inclusion problem $L(A)\subseteq L(B)$ is PSPACE-hard for arbitrary 
B\"uchi automata and therefore unlikely to be definable in $\hmu{\omega}$. In the 
restricted case of $B$ being deterministic, it becomes 
solvable in polynomial time. Remember that a B\"uchi automaton is called deterministic if
for all $a\in \Sigma$, for all $s,s_1,s_2\in S$, if $\trans s a {s_1}$ and 
$\trans s a {s_2}$, then 
$s_1=s_2$. 

Let us now encode the language inclusion problem $L(A)\subseteq L(B)$ as
a $\hmu{\omega}$ model checking problem. To shorten a bit the formula,
we assume that $B$ is moreover \emph{complete}, \emph{i.e.} 
for all $s\in S$, for all 
$a\in \Sigma$, there is at least one $s'$ such that $\trans s a {s'}$.
Let us introduce the modality 
$\Mudiam{synch}{}\varphi==\bigvee_{a\in \Sigma}\Mudiam{a}{x}\Mudiam{a}{y}\varphi$.
Consider the formula
$$
\phi_{incl}~~==~~\Mudiam{synch}{}^*\nu Z_1.\Big(\prdc{final}(x) /| \neg\prdc{final}(y) /| 
\mu Z_2. \Mudiam{synch}{} \big(Z_1|/(\neg \prdc{final}(y)/| Z_2)\big)\Big)
$$
Let $\lts_{A,B}$ be the LTS obtained as the disjoint union of $A$ and $B$ with 
initial states $s_A$ of $A$ and $s_B$ of $B$ respectively.
Then $L(A)$ is included in $L(B)$ if and only if 
$\lts_{A,B}, v \not|= \phi_{incl}$ where $v(x) = s_A$ and $v(y) = s_B$.
Indeed, this formula is satisfied
if there is a
run $r_A$ of $A$ and a run $r_B$ of $B$ reading the same word 
$w\in \Sigma^\omega$ such
that $r_A$ visits a final state of $A$ infinitely often, whereas $r_B$
eventually stops visiting the final states of $B$. Since $B$ is deterministic,
no other run $r_B'$ could read $w$, thus $w\in L(A)\backslash L(B)$.

The same ideas can be applied to parity automata. A parity automaton 
is a finite automaton where states are assigned priorities; it can be seen
as an LTS $(S,s_0,@D,\rho)$ where $\rho$ interprets 
\emph{priority predicates} $\prdc{prty}_k$ in such a way that 
$\rho(s)$ is a singleton $\{\prdc{prty}_k\}$ for all $s\in S$.
A word $w\in \Sigma^{\omega}$ is accepted by a parity automaton
if there is a run of $w$ such that
the largest priority visited infinitely often is even. 
Consider the formulas 
$\prdc{prty}_{\leq m}(x)=\prdc{prty}_0(x)|/\dots|/\prdc{prty}_m(x)$ and
$$
\phi_{n,m} ~~=~~ \Mudiam{synch}{}^*\nu Z.\Mudiam{synch'}{}^+\big(\prdc{prty}_n(x) /| 
\Mudiam{synch'}{}^+(\prdc{prty}_m(y) /|Z)\big) 
$$
where $\Mudiam{synch'}{}^+\phi$ is a shorthand for
$ \mu Z.\Mudiam{synch}{}
\prdc{prty}_{\leq n}(x)/|\prdc{prty}_{\leq m}(y)/| (\phi |/ Z)
$.
Then $\phi_{n,m}$ asserts that there are
two runs $r_A$ and $r_B$ 
of two parity automata $A$ and $B$ recognizing the same word $w$ such that 
the highest priorities 
visited infinitely often by $r_A$ 
and $r_B$ are respectively $n$ and $m$. Since
$L(A)\not\subseteq L(B)$ if and only there is an even $n$ and an odd $m$ such
that $\lts_{A,B}\models\phi_{n,m}$, this gives us again a decision procedure
for the language inclusion problem of parity automata 
when $B$ is deterministic complete.

\paragraph*{Parsing of Formal Languages.}
A third application of $\hmu{\omega}$ is in the field of parsing for formal,
namely context-free languages.
To each finite word $w$, we may associate its linear LTS $\lts_w$. For 
instance, for $w=aab$, $\lts_{w}$ is the LTS
{\small
\begin{tikzpicture}[baseline]
  \matrix[column sep=.7cm,nodes={shape=circle,draw,minimum size=1.5mm}] {
    \node (1) {}; & \node (2) {} ; & \node (3) {}; & \node (4) {}; \\
  };
  \draw[->] (1) -- node[above] {$a$} (2);
  \draw[->] (2) -- node[above] {$a$} (3);
  \draw[->] (3) -- node[above] {$b$} (4);
\end{tikzpicture}
}.
Let us now consider a context-free grammar $G$, and define a formula 
that describes the language of $G$. To ease the presentation, 
we assume that $G$ is 
in Chomsky normal form, but a linear-size formula would be derivable 
for an arbitrary context-free grammar as well. The production rules of $G$ are thus
of the form
either $X_i->X_jX_k$ or $X_i->a$, for $X_1,\dots,X_n$ the non-terminals of $G$. 
Let us pick variables $x$,$y$ and $z$, intended to represent 
respectively 
the initial the final, and an intermediate position in 
the (sub)word currently parsed.
To every non-terminal $X_i$, we associate the recursive definition:
$$
\phi_i ~~=_{\mu}~~ \bigvee_{X_i->a} \Mudiam{a}{x}~x\sim y ~~|/ ~~
\bigvee_{X_i->X_jX_k} \repl{z<-x}\Mudiam{-}{z}^*\big((\repl{y<-z}\phi_j)/|(\repl{x<-z}\phi_k)\big) 
$$
where $x\sim y$ is the formula characterizing bisimilarity and 
$\Mudiam{-}{z}^*\phi$ is $\mu Z.\phi |/\bigvee_{a\in \Sigma}\Mudiam{a}{z}Z$.
If $v(x)$ and $v(y)$ are respectively the initial and final states of $\lts_w$,
then $\lts_w,v|=\phi_i$ is equivalent to $w$ being derivable in $G$ starting 
with the symbol $X_i$.

\paragraph*{String Problems.}
Model Checking for $\hmu{\infty}$ can even be useful for computation (as opposed
to decision) problems. Consider for example the Longest Common Subword problem:
given words $w_1,\ldots,w_m$ over some alphabet $\Sigma$, find a longest $v$ that
is a subword of all $w_i$. This problem is NP-complete for an unbounded number of
input words. Thus, we consider the problem restricted to some fixed $m$, and it is
possible to define a formula $\varphi^m_{\mathrm{LCSW}} \in \hmu{m}$ such that model
checking this formula on a suitable representation of the $w_i$ essentially 
computes such a common subword.

For the LTS take the disjoint union of all $\lts_{w_i}$ for $i=1,\ldots,m$, and
assume that each state in $\lts_{w_i}$ is labeled with a proposition $p_i$ which
makes it possible to define $m$-tuples of states in which the $i$-th component
belongs to $\lts_{w_i}$. Now consider the formula
\begin{displaymath}
\varphi^m_{\mathrm{LCSW}} := 
\nu X.\bigwedge\limits_{i=1}^m p_i(x_i) \wedge \bigvee\limits_{a \in \Sigma} 
\Mudiam{a}{1}\ldots\Mudiam{a}{m} X
\end{displaymath}
Note that $\varphi^m_{\mathrm{LCSW}}$ is unsatisfiable for any $m \ge 1$. Thus,
a symbolic model checking algorithm 
for instance
would always return the empty set of tuples when called on this formula and any LTS.
However, on an LTS representing $w_1,\ldots,w_m$ as described above it consecutively
computes in the $j$-th round of the fixpoint iteration, all tuples of positions
$h_1,\ldots,h_m$ such that the subwords in $w_i$ from position $h_i-j$ to $h_i$ are 
all the same for every $i=1,\ldots,m$. Thus, it computes, in its penultimate round
the positions inside the input words in which the longest common substrings end. Their
starting points can easily be computed by maintaining a counter for the number of
fixpoint iterations done in the model checking run.

In the same way, it is possible to compute the longest common subsequence of input
words $w_1,\ldots,w_m$. A subsequence of $w$ is obtained by deleting arbitrary symbols,
whereas a subword is obtained by deleting an arbitrary prefix and suffix from $w$. The
Longest Common Subsequence problem is equally known to be NP-complete for unbounded
$m$. For any fixed $m$, however, the following formula can be used to compute all
longest common subsequences of such input words using model checking technology in the 
same way as it is done in the case of the Longest Common Subword problem.
\begin{displaymath}
\varphi^m_{\mathrm{LCSS}} := 
\nu X.\bigwedge\limits_{i=1}^m p_i(x_i) \wedge \bigvee\limits_{a \in \Sigma} 
\Mudiam{a}{x_1}\Mudiam{-}{x_1}^*\ldots\Mudiam{a}{x_m}\Mudiam{-}{x_m}^* X
\end{displaymath}
where $\Mudiam{-}{x_i}^*\psi$ stands for $\mu Y.\psi \vee \bigvee\limits_{a \in \Sigma} \Mudiam{a}{x_i} Y$.

\paragraph*{Games.}

The Cat and Mouse Game is played on a directed graph with three
distinct nodes $c$, $m$ and $t$ as follows. Initially, the cat resides in node $c$,
the mouse in node $m$. In each round, the mouse moves first. He can move along an 
edge to a successor node of the current one or stay on the current node, then the 
cat can do the same. If the cat reaches the mouse, she wins; otherwise, if the mouse reaches the target node $t$, he wins; otherwise,
the mouse
runs forever without being caught nor reaching the target node: in 
that case, the cat wins.
The problem of solving the Cat and Mouse Game is to decide whether or 
not the mouse has a winning strategy for a given graph. 

Note that this problem is not bisimulation-invariant under the straight-forward encoding
of the directed graph as an LTS with a single proposition $t$ to mark the target node.
Consider for example the following two, bisimilar game arenas.
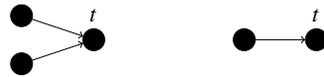
\begin{center}
\scalebox{0.8}{\begin{tikzpicture}[minimum size=3mm]
  \node[state, minimum size=3mm, fill] (n0) {};
  \node (a0) [right of=n0, node distance=12mm] {};
  \node[state, minimum size=3mm, fill] (n1) [below of=n0, node distance=8mm] {};
  \node[state, minimum size=3mm, fill] (n2) [below of=a0, node distance=4mm] {};

  \node (l2) [above of=n2, node distance=4mm] {$t$};

  \node[state, minimum size=3mm, fill] (n3) [right of=n2, node distance=2.5cm] {};
  \node[state, minimum size=3mm, fill] (n4) [right of=n3, node distance=12mm] {};
  \node (l4) [above of=n4, node distance=4mm] {$t$};

  \path[->] (n0) edge (n2)
            (n1) edge (n2)
            (n3) edge (n4);

\end{tikzpicture}}
\end{center} 
Clearly, if the cat and mouse start on the two separate leftmost nodes then the mouse
can reach the target first. However, these nodes are bisimilar to the left 
node of the
right graph, and if they both start on this one then the cat has caught the mouse
immediately. 

Thus, winning strategies cannot necessarily be defined in $\hmu{\infty}$. However, it
is possible to define them when a new atomic formula $\mathit{eq}(x,y)$ expressing
that $x$ and $y$ evaluate to the same node, is being added to the syntax of 
$\hmu{\infty}$ (standard model checking procedures can
be extended to handle the equality predicate $\mathit{eq}$ as
well).

\begin{displaymath}
\varphi_{\mathrm{CMG}} := \mu X.(t(x) \wedge \neg\mathit{eq}(x,y)) \vee 
\Mudiam{-}{x}(\neg\mathit{eq}(x,y)) \wedge \Mubox{-}{y} X)
\end{displaymath}
We have $v \models \varphi_{\mathrm{CMG}}$ if and only if 
the mouse can win from position
$v(x)$ when the cat is on position $v(y)$ initially.



\section{Conclusion}
\vskip-2mm
We have considered the modal fixpoint logic $\hmu{\omega}$ for a potential use
in algorithm design and given examples of problems which can be defined
in $\hmu{\omega}$. The combination of fixpoint quantifiers and modal operators
has been proved to be very fruitful for obtaining algorithmic solutions for problems in automatic
program verification. The examples boost the idea of using successful model checking technology
in other areas too. 

The use of model checking algorithms on fixed formulas does not provide a generic recipe that 
miraculously generates efficient algorithms, but it provides the potential to do so. The next step
on this route towards an efficient algorithm for some problem $P$ requires partial evaluation on a 
model checking algorithm and the formula $\varphi_P$ defining $P$. This usually requires manual tweaking
of the algorithm and is highly dependent on the actual $\varphi_P$. Thus, future work on this direction
would consist of consequently optimising $\hmu{\omega}$ model checking algorithms for certain definable
problems and testing their efficiency in practice.

On a different note, $\hmu{\omega}$ is an interesting fixpoint calculus for which the model checking problem over infinite-state transition 
systems has not been quite studied so far. The most prominent result in this area is the decidability of $\hmu{1}$ over pushdown 
LTS~\cite{Walukiewicz96}. However, model checking $\hmu{\omega}$ --- or even just $\hmu{k}$ for some 
$k \ge 2$ --- seems undecidable for pushdown LTS. It is questionable whether model checking of 
$\hmu{\omega}$ is decidable for any popular class of infinite-state transition systems.
\vspace*{-3mm}


\bibliographystyle{eptcs}
\bibliography{biblio}

\end{document}